\documentstyle[twoside]{article}

\catcode`\@=11
\long\def\@makefntext#1{
\protect\noindent \hbox to 3.2pt {\hskip-.9pt  
$^{{\eightrm\@thefnmark}}$\hfil}#1\hfill}		%CAN BE USED 

\def\thefootnote{\fnsymbol{footnote}}
\def\@makefnmark{\hbox to 0pt{$^{\@thefnmark}$\hss}}	%ORIGINAL 
	
\def\ps@myheadings{\let\@mkboth\@gobbletwo
\def\@oddhead{\hbox{}
\rightmark\hfil\eightrm\thepage}   
\def\@oddfoot{}\def\@evenhead{\eightrm\thepage\hfil
\leftmark\hbox{}}\def\@evenfoot{}
\def\sectionmark##1{}\def\subsectionmark##1{}}

\oddsidemargin=\evensidemargin
\renewcommand{\thefootnote}{\fnsymbol{footnote}}

\newcounter{sectionc}\newcounter{subsectionc}\newcounter{subsubsectionc}
\renewcommand{\section}[1] {\vspace{12pt}\addtocounter{sectionc}{1} 
\setcounter{subsectionc}{0}\setcounter{subsubsectionc}{0}\noindent 
	{\tenbf\thesectionc. #1}\par\vspace{5pt}}
\renewcommand{\subsection}[1] {\vspace{12pt}\addtocounter{subsectionc}{1} 
	\setcounter{subsubsectionc}{0}\noindent 
	{\bf\thesectionc.\thesubsectionc. {\kern1pt \bfit #1}}\par\vspace{5pt}}
\renewcommand{\subsubsection}[1] {\vspace{12pt}\addtocounter{subsubsectionc}{1}
	\noindent{\tenrm\thesectionc.\thesubsectionc.\thesubsubsectionc.
	{\kern1pt \tenit #1}}\par\vspace{5pt}}
\newcommand{\nonumsection}[1] {\vspace{12pt}\noindent{\tenbf #1}
	\par\vspace{5pt}}

\newcounter{appendixc}
\newcounter{subappendixc}[appendixc]
\newcounter{subsubappendixc}[subappendixc]
\renewcommand{\thesubappendixc}{\Alph{appendixc}.\arabic{subappendixc}}
\renewcommand{\thesubsubappendixc}
	{\Alph{appendixc}.\arabic{subappendixc}.\arabic{subsubappendixc}}

\renewcommand{\appendix}[1] {\vspace{12pt}
        \refstepcounter{appendixc}
        \setcounter{figure}{0}
        \setcounter{table}{0}
        \setcounter{lemma}{0}
        \setcounter{theorem}{0}
        \setcounter{corollary}{0}
        \setcounter{definition}{0}
        \setcounter{equation}{0}
        \renewcommand{\thefigure}{\Alph{appendixc}.\arabic{figure}}
        \renewcommand{\thetable}{\Alph{appendixc}.\arabic{table}}
        \renewcommand{\theappendixc}{\Alph{appendixc}}
        \renewcommand{\thelemma}{\Alph{appendixc}.\arabic{lemma}}
        \renewcommand{\thetheorem}{\Alph{appendixc}.\arabic{theorem}}
        \renewcommand{\thedefinition}{\Alph{appendixc}.\arabic{definition}}
        \renewcommand{\thecorollary}{\Alph{appendixc}.\arabic{corollary}}
        \renewcommand{\theequation}{\Alph{appendixc}.\arabic{equation}}
        \noindent{\tenbf Appendix \theappendixc #1}\par\vspace{5pt}}
\newcommand{\subappendix}[1] {\vspace{12pt}
        \refstepcounter{subappendixc}
        \noindent{\bf Appendix \thesubappendixc. {\kern1pt \bfit #1}}
	\par\vspace{5pt}}
\newcommand{\subsubappendix}[1] {\vspace{12pt}
        \refstepcounter{subsubappendixc}
        \noindent{\rm Appendix \thesubsubappendixc. {\kern1pt \tenit #1}}
	\par\vspace{5pt}}

\newcommand{\textlineskip}{\baselineskip=13pt}
\newcommand{\smalllineskip}{\baselineskip=10pt}

\def\eightcirc{
\begin{picture}(0,0)
\put(4.4,1.8){\circle{6.5}}
\end{picture}}
\def\eightcopyright{\eightcirc\kern2.7pt\hbox{\eightrm c}}

\def\abstracts#1#2#3{{
	\centering{\begin{minipage}{4.5in}\baselineskip=10pt\footnotesize
	\parindent=0pt #1\par 
	\parindent=15pt #2\par
	\parindent=15pt #3
	\end{minipage}}\par}} 

\newcommand{\bibit}{\nineit}

\renewenvironment{thebibliography}[1]
	{\frenchspacing
	 \ninerm\baselineskip=11pt
	 \begin{list}{\arabic{enumi}.}
	{\usecounter{enumi}\setlength{\parsep}{0pt}
	 \setlength{\leftmargin 12.7pt}{\rightmargin 0pt} %FOR 1--9 ITEMS
	 \setlength{\itemsep}{0pt} \settowidth
	{\labelwidth}{#1.}\sloppy}}{\end{list}}

\newcounter{itemlistc}
\newcounter{romanlistc}
\newcounter{alphlistc}
\newcounter{arabiclistc}

\newcommand{\fcaption}[1]{
        \refstepcounter{figure}
        \setbox\@tempboxa = \hbox{\footnotesize Fig.~\thefigure. #1}
        \ifdim \wd\@tempboxa > 5in
           {\begin{center}
        \parbox{5in}{\footnotesize\smalllineskip Fig.~\thefigure. #1}
            \end{center}}
        \else
             {\begin{center}
             {\footnotesize Fig.~\thefigure. #1}
              \end{center}}
        \fi}

\newcommand{\tcaption}[1]{
        \refstepcounter{table}
        \setbox\@tempboxa = \hbox{\footnotesize Table~\thetable. #1}
        \ifdim \wd\@tempboxa > 5in
           {\begin{center}
        \parbox{5in}{\footnotesize\smalllineskip Table~\thetable. #1}
            \end{center}}
        \else
             {\begin{center}
             {\footnotesize Table~\thetable. #1}
              \end{center}}
        \fi}

\def\@citex[#1]#2{\if@filesw\immediate\write\@auxout
	{\string\citation{#2}}\fi
\def\@citea{}\@cite{\@for\@citeb:=#2\do
	{\@citea\def\@citea{,}\@ifundefined
	{b@\@citeb}{{\bf ?}\@warning
	{Citation `\@citeb' on page \thepage \space undefined}}
	{\csname b@\@citeb\endcsname}}}{#1}}

\newif\if@cghi
\def\cite{\@cghitrue\@ifnextchar [{\@tempswatrue
	\@citex}{\@tempswafalse\@citex[]}}
\def\citelow{\@cghifalse\@ifnextchar [{\@tempswatrue
	\@citex}{\@tempswafalse\@citex[]}}
\def\@cite#1#2{{$\null^{#1}$\if@tempswa\typeout
	{IJCGA warning: optional citation argument 
	ignored: `#2'} \fi}}

\def\pmb#1{\setbox0=\hbox{#1}
	\kern-.025em\copy0\kern-\wd0
	\kern.05em\copy0\kern-\wd0
	\kern-.025em\raise.0433em\box0}

\def\fnt#1#2{\footnotetext{\kern-.3em
	{$^{\mbox{\scriptsize #1}}$}{#2}}}

\def\fpage#1{\begingroup
\voffset=.3in
\thispagestyle{empty}\begin{table}[b]\centerline{\footnotesize #1}
	\end{table}\endgroup}

\def\runninghead#1#2{\pagestyle{myheadings}
\markboth{{\protect\footnotesize\it{\quad #1}}\hfill}
{\hfill{\protect\footnotesize\it{#2\quad}}}}
\font\tenrm=cmr10
\font\tenit=cmti10 
\font\tenbf=cmbx10
\font\bfit=cmbxti10 at 10pt
\font\ninerm=cmr9
\font\nineit=cmti9

\font\eightrm=cmr8

\def\qed{\hbox{${\vcenter{\vbox{			%HOLLOW SQUARE
   \hrule height 0.4pt\hbox{\vrule width 0.4pt height 6pt
   \kern5pt\vrule width 0.4pt}\hrule height 0.4pt}}}$}}

\renewcommand{\thefootnote}{\fnsymbol{footnote}}	%USE SYMBOLIC FOOTNOTE

\begin{document}

\runninghead{The Spectrum of Diquark Composites in Cold Dense QCD} 
{The Spectrum of Diquark Composites in Cold Dense QCD}

\normalsize\textlineskip
\thispagestyle{empty}
\setcounter{page}{1}

%\copyrightheading{}		%{Vol. 0, No. 0 (1993) 000--000}

%\vspace*{0.88truein}

\fpage{1}
\centerline{\bf THE SPECTRUM OF DIQUARK COMPOSITES}
\vspace*{0.035truein}
\centerline{\bf IN COLD DENSE QCD}
\vspace*{0.37truein}
\centerline{\footnotesize IGOR SHOVKOVY\footnote{On leave of 
                     absence from
                     Bogolyubov Institute for Theoretical
                     Physics, 252143, Kiev, Ukraine.} 
\footnote{Present
address: School of Physics and Astronomy, University of Minnesota,
         Minneapolis, MN 55455, USA.}}
\vspace*{0.015truein}
\centerline{\footnotesize\it Physics Department, University
of Cincinnati}
\baselineskip=10pt
\centerline{\footnotesize\it Cincinnati, Ohio 45221-0011, 
USA}
\vspace*{0.225truein}
%\publisher{(received date)}{(revised date)}

\vspace*{0.21truein}
\abstracts{The Bethe-Salpeter equations for spin zero diquark composites
in the color superconducting phase of $N_f=2$ and $N_f=3$ cold dense QCD
are studied. The explicit form of the spectrum of the diquarks with the
quantum numbers of the (pseudo-) Nambu-Goldstone bosons is derived.}{}{}

\textlineskip			%) USE THIS MEASUREMENT WHEN THERE IS
\vspace*{12pt}			%) NO SECTION HEADING
%
%\vspace*{1pt}\textlineskip	%) USE THIS MEASUREMENT WHEN THERE IS
%\section{General Appearance}	%) A SECTION HEADING
%\vspace*{-0.5pt}
\noindent
In this talk, I report on the progress in studying the spectrum of diquark
bound states in color superconducting phase of cold dense QCD. It is based
on a series of papers done in collaboration with V.A.~Miransky and
L.C.R.~Wijewardhana.\cite{bs-all}

Despite recent advances in studying the color superconducting phase of
dense QCD,\cite{first,SPR,SD,SonSt&Others} the detailed spectrum of
diquark bound states (mesons) in such a phase is still poorly known. In
fact, most of the existing studies deal only with the case of the
Nambu-Goldstone (NG) bosons.\cite{SonSt&Others} This is partly because
many properties of such states could be derived from symmetry arguments
alone.

In order to determine the spectrum of other diquark bound states, one
should consider the Bethe-Salpeter (BS) equations in the appropriate
channels.\cite{bs-all} Of special interest are the bound states with the
same quantum numbers as those of the NG bosons. Indeed, recently we
argued\cite{us3} that the spectrum of a color superconductor may contain
an infinite tower of massive diquark states with such quantum numbers.
 
Let me start with the case of two flavor QCD. The original $SU(3)_{c}$
color gauge group is broken down to $SU(2)_{c}$ subgroup by the Higgs
mechanism. Five NG bosons [a doublet, an antidoublet and a singlet with
respect to the unbroken $SU(2)_{c}$] are removed from the physical
spectrum, giving masses to five gluons. The spectrum contains, however,
five (nearly) massless pseudo-NG bosons, and an infinite tower of massive
diquark singlets [with respect to the unbroken $SU(2)_{c}$ subgroup].
Furthermore, the following leading order mass formula is derived for the
singlets:
\begin{equation}
M^{2}_{n} \simeq 4 |\Delta |^{2}
\left(1-\frac{\alpha_{s}^{2}\kappa}{(2n+1)^{4}}\right),
\quad n=1,2,\dots,
\label{mass-singlet}
\end{equation}
\pagebreak
\textheight=7.8truein
\setcounter{footnote}{0}
\renewcommand{\thefootnote}{\alph{footnote}}

\noindent
where $\kappa$ is a constant of order 1, $|\Delta |$ is the dynamical
Majorana mass of quarks, and $\alpha_{s}$ is the value of the running
coupling constant related to the scale of the chemical potential $\mu$.

Some qualitative features of the diquark pairing dynamics at large $\mu$
could be understood even without a detailed analysis of the BS equations.
The important point to notice is that the left- and right-handed sectors
of the theory are approximately decoupled. Then, as far as it concerns the
diquark paring dynamics, the leading order of QCD would not change if the
gauge group is enlarged from $SU(3)_{c}$ to the approximate
$SU(3)_{c,L}\times SU(3)_{c,R}$. In the theory with $SU(3)_{c,L}\times
SU(3)_{c,R}$ gauge group, the pattern of the symmetry breaking is
$SU(3)_{c,L}\times SU(3)_{c,R} \to SU(2)_{c,L} \times SU(2)_{c,R}$, and
ten NG bosons should appear. They are, however, unphysical because of the
Higgs mechanism. Since the true gauge group of QCD is vector-like
$SU(3)_{c}$, only five NG bosons (scalars) are removed from the spectrum
of physical particles. The other five NG bosons (pseudoscalars) should
remain in the spectrum. Clearly, in the complete theory, these latter are
the pseudo-NG bosons, getting non-zero masses due to higher order
(possibly, non-perturbative) corrections. At large $\mu$, it is natural to
expect that the masses of the pseudo-NG bosons are small even compared to
the value of the dynamical quark mass.

While omitting the detailed analysis of the BS equations,\cite{bs-all} let
me mention that the structure of the BS wave functions in the case of the
(pseudo-) NG bosons is unambiguously obtained from the Ward identities. By
making use of the solution, it is straightforward to derive the decay
constants and the velocities of the (pseudo-) NG bosons in the
Pagels-Stokar approximation. The result reads\cite{bs-all}
\begin{equation}  
F=\frac{\mu}{2\sqrt{2}\pi} \qquad \mbox{and} \qquad 
v=\frac{1}{\sqrt{3}}.
\label{dec}
\end{equation}
Now, what is the role of the Meissner effect in the problem at hand? As 
is well known, this effect is irrelevant for the dynamics of the gap
formation in the leading order of dense QCD.\cite{SD} The paring dynamics
of the massive diquarks, on the other hand, is very sensitive to the
Meissner effect.\cite{bs-all} This is because of the quasiclassical nature
of the corresponding bound states whose binding energy is small compared
to the value of the quark mass. The detailed analysis shows, in fact, that
only the gluons of the unbroken $SU(2)_{c}$ could provide a strong enough
attraction to form massive radial excitations of the (pseudo-) NG bosons.
Essentially, this is the reason why there are no towers of massive states
in the doublet and antidoublet channels.

Because of the approximate degeneracy between the left- and right-handed
(or, equivalently, scalar and pseudoscalar) sectors of dense QCD, the
spectrum of massive states reveals the property of parity doubling. This
means, in particular, that all diquark singlets with masses in
Eq.~(\ref{mass-singlet}) come in pairs of approximately degenerate
parity even and parity odd states.

At the end, let me briefly describe the situation in dense QCD with three
quark flavors. The ground state of this model is the so-called
color-flavor locked (CFL)  phase.\cite{CFL} The original gauge symmetry
$SU(3)_{c}$ and the global chiral symmetry $SU(3)_{L} \times SU(3)_{R}$
break down to the global diagonal $SU(3)_{c+L+R}$ subgroup. Out of total
sixteen (would be) NG bosons, eight are removed from the physical spectrum
by the Higgs mechanism, providing masses to eight gluons. The other eight
NG bosons show up as an octet [under the unbroken $SU(3)_{c+L+R}$] of
physical particles. In addition, the global baryon number symmetry and the
approximate $U(1)_{A}$ symmetry also get broken, and an extra NG boson and
a pseudo-NG boson appear in the low energy spectrum. They both are
singlets under $SU(3)_{c+L+R}$.

The analysis of the BS equations is similar to that in the two flavor
case. The structure of the BS wave functions of the (pseudo-) NG bosons is
again derived from the Ward identities. The Pagels-Stokar approximation
gives qualitatively the same results for the decay constants and the
velocities as in Eq.~(\ref{dec}). In $N_f=3$ dense QCD, however, there are
no massive radial excitations of the (pseudo-) NG bosons. In brief, this
is related to the fact that all eight gluons are subject to the Meissner
screening in the CFL phase.\cite{bs-all}

In conclusion, the spectrum of diquarks with the quantum numbers of the
(pseudo-) NG bosons is determined by using the method of the BS equation.
The existance of five pseudo-NG bosons in $N_{f}=2$ dense QCD is revealed
for the first time. Taking these diquarks into account may be important
for deriving some thermodynamic properties of dense QCD.

\nonumsection{Acknowledgements} 
\noindent 
I would like to thank V.~Miransky for useful comments on the manuscript.
This work was supported by the U.S. Department of Energy Grant
No.~DE-FG02-84ER40153.

\nonumsection{References}


\noindent
\begin{thebibliography}{000}

\bibitem{bs-all}
V. A. Miransky, I. A. Shovkovy, and L. C. R. Wijewardhana, 
hep-ph/0003327; 
Phys. Rev. D {\bf 62}, 085025 (2000); 
hep-ph/0009173.

\bibitem{first} M.~Alford, K.~Rajagopal, and F.~Wilczek,
Phys. Lett. B{\bf 422}, 247 (1998);
R.~Rapp, T.~Sch\"{a}fer, E.V.~Shuryak, and M.~Velkovsky, 
Phys. Rev. Lett. {\bf 81}, 53 (1998).

\bibitem{SPR} 
D. T. Son, Phys. Rev. D {\bf 59}, 094019 (1999);
R. D. Pisarski and D. H. Rischke, Phys. Rev. Lett. {\bf 83}, 37 (1999).

\bibitem{SD} 
T. Schafer and F. Wilczek, Phys. Rev. D {\bf 60}, 114033 (1999);
D. K. Hong, V. A. Miransky, I. A. Shovkovy and L. C. R. Wijewardhana, 
    {\bibit ibid.} {\bf 61}, 056001 (2000);
R. D. Pisarski and D. H. Rischke, 
    {\bibit ibid.} {\bf 61}, 051501 (2000);
W. E. Brown, J. T. Liu and H.-C. Ren,
    {\bibit ibid.} {\bf 61}, 114012 (2000);
S. D. H. Hsu and M. Schwetz,
    Nucl. Phys. {\bf B572}, 211 (2000);
I. A. Shovkovy and L. C. R. Wijewardhana,
    Phys. Lett. B {\bf 470}, 189 (1999); 
T. Schafer, Nucl. Phys. {\bf B575}, 269 (2000).


\bibitem{SonSt&Others}
D. T.~Son and M. A. Stephanov,
    Phys. Rev. D {\bf 61}, 074012 (2000);
    {\bibit ibid.} {\bf 62}, 059902(E) (2000);
K.~Zarembo, {\bibit ibid.} {\bf 62}, 054003 (2000);
R.~Casalbuoni and R.~Gatto,
    Phys. Lett. B {\bf 464}, 111 (1999);
M.~Rho, A.~Wirzba and I.~Zahed,
    {\bibit ibid.} {\bf 473}, 126 (2000);
D.K.~Hong, T.~Lee and D.-P.~Min,
    {\bibit ibid.} {\bf 477}, 137 (2000);
C.~Manuel and M.G.H.~Tytgat, 
    {\bibit ibid.} {\bf 479}, 190 (2000);
S.R.~Beane, P.F.~Bedaque and M.J.~Savage,
    {\bibit ibid.} {\bf 483}, 131 (2000);
M.~Rho, E.~Shuryak, A.~Wirzba and I.~Zahed,
    Nucl. Phys. {\bf A676}, 273 (2000).

\bibitem{us3}
V. A. Miransky, I. A. Shovkovy and L. C. R. Wijewardhana,
Phys. Lett. B {\bf 468}, 270 (1999).

\bibitem{CFL}
M.~Alford, K.~Rajagopal, and F.~Wilczek,
    Nucl. Phys. {\bf B537}, 443 (1999).

\end{thebibliography}
\end{document}